\documentclass[secnumarabic,aps,prd,floatfix,nofootinbib,showkeys,twocolumn,showpacs,10pt,superscriptaddress]
{revtex4-1}
\usepackage{epsfig}
\usepackage{amsmath}

\begin{document}
\begin{titlepage}

\begin{center}
{\Large \bf Thermodynamic Equilibrium in General Relativity
 }

\vskip 0.1cm

J. A. S. Lima$^{a,}$\footnote{jas.lima@iag.usp.br}, A. Del Popolo$^{b,c,d,}$\footnote{adelpopolo@oact.inaf.it}, A. R. Plastino$^{e,}$\footnote{arplastino@unnoba.edu.ar}
\end{center}

\begin{quote}
$^a$Departamento de Astronomia, Universidade de S\~ao Paulo, Rua do
Mat\~ao 1226,
05508-900, S\~ao Paulo, SP, Brazil\\
$^b$Dipartimento di Fisica e Astronomia, Catania University, Via S. Sofia, 64, 95123 Catania, Italy\\
$^c$Institute of Astronomy, Russian Academy of Sciences, Pyatnitskaya str., 48, Moscow 119017, Russia\\
$^d$INFN sezione di Catania, Catania, Italy\\
$^e$CeBio y Departamento de Ciencias B\'asicas,
Universidad Nacional del Noroeste de la Provincia de Buenos Aires,
UNNOBA, Conicet, Roque Saenz Pena 456, Junin, Argentina\\
\end{quote}

\centerline{\bf ABSTRACT}
\bigskip

The thermodynamic equilibrium condition for a static self-gravitating fluid in the Einstein theory is defined by the Tolman-Ehrenfest temperature law,  $T{\sqrt {g_{00}(x^{i})}} = constant$,  according to which the proper temperature depends explicitly on the position within the medium through the metric coefficient $g_{00}(x^{i})$. By assuming the validity of Tolman-Ehrenfest ``pocket temperature'', Klein also proved a similar relation for the chemical potential, namely, $\mu {\sqrt {g_{00}(x^{i})}} = constant$. In this letter we prove that a more general relation uniting both quantities holds regardless of the equation of state satisfied by the medium, and
that the original Tolman-Ehrenfest law form is valid only if the chemical potential vanishes identically. In the general case of equilibrium, the temperature and the chemical potential are intertwined in such a way that only a
definite (position dependent) relation uniting both quantities is obeyed.  As an illustration of these results, the temperature expressions for an isothermal gas (finite spherical distribution) and a neutron star are also determined.

\end{titlepage}
\pagestyle{plain} 

\vskip 0.3cm

\section{Introduction}

It is usually believed, at least for non members of the general relativity community, that equality of temperature is a condition for thermal equilibrium between two systems or between two parts of a single system (``Zeroth" law of thermodynamics). Furthermore, the second law of thermodynamics, in one of its variants, e.g. Clausius' one, states that heat flows from a hotter to a colder medium, till thermal equilibrium is finally restored. However, these both basic conditions can be  violated in the framework of general relativity.

Many decades ago, Tolman and Ehrenfest discussed how to determine
the temperature distribution within a self-gravitating fluid that
has come to thermal equilibrium. The result was a remarkable thermo-gravitational effect in the
framework of general relativity: heat, as any other source of energy, is subjected to gravity.
The preliminary results assuming spherical symmetry
were obtained by Tolman {in the weak field approximation} \cite{T30}, but a proof of the theorem valid
for a more general static field was published in a subsequent
paper by Tolman and Ehrenfest \cite{TE30}(see also \cite{Tolman34}). In order to discuss
the equilibrium temperature to this particular case, they assumed
that the self-gravitating fluid generates a static gravitational
field described by the line element
\begin{equation}
\label{M} ds^2=g_{00}c^{2}dt^2 - g_{ij}dx^{i}dx^{j}\,,
\end{equation}
where Latin indices denote spatial coordinates {and the signature adopted here is} $(+,-,-,-)$. The components
$g_{00}(x^{r})$ and $g_{ij}(x^{r})$ are independent of time but
depends in an arbitrary way of the spatial coordinates $x^{r}\,\,
(r=1,2,3)$.  Under such conditions, the ``pocket temperature" Tolman-Ehrenfest (TE) theorem can
be expressed as

\begin{equation}
 T{\sqrt g_{00}(x^{i})}= \tilde{T}=constant\,,
\end{equation} where $\tilde{T}$ is constant in all parts of the system.
The interesting aspect of this relation is that the proper
temperature necessarily varies from point to point within the
self-gravitating fluid that has come to equilibrium, thereby violating the
so-called ``Zeroth" law of thermodynamics \cite{S93}. This
result is nowadays considered an important key in the framework of
black-hole physics\cite{BH}, or more generally, to compact objects in the astrophysical domain.
Tolman stressed that the temperature $T$ is directly measurable by local
observers, and as such, it must be considered the fundamental
quantity that we mean by temperature at a given point.

In principle, due to its physical interest, the Tolman
distribution law demands a more closer scrutiny. As it appears,
the proof of the TE theorem is very particular in many different
aspects. To begin with, since the self-gravitating fluid may have
a generic nature, they first assumed that the parts of the system
whose temperature need to be compared are in thermal contact with
a small connecting tube containing blackbody radiation, or at
least could be put in such contacting device without perturbing
the system which should be considered a kind of reservoir. In other
words, the tube works like a radiation thermometer. Second, the
energy conservation law was applied to the thermometer itself not
to the fluid source of curvature as it should be desirable in
principle. Finally, it should be remarked that black body
radiation is a very special kind of medium since its chemical
potential is zero, and its basic thermodynamic quantities
(temperature, pressure and energy density) are related in a very
simple way. Moreover, photons suffers gravitational redshift in
the presence of a static field, a kinematic phenomenon that is not
directly related with the idea of thermal equilibrium.

In literature, several have been the trials to determine or extend the { TE theorem}. For example Buchdahl \citep{Buchdahl1949} formally extended TE result to {self-gravitating fluids supporting stationary spacetime} through the time-like killing vector
\begin{equation}
K^a= (\partial_t)^a=(1,0,0,0)^a
\end{equation}
As shown by \citep{Santiago2018},  even considering that the approach of \citep{Buchdahl1949} looks similar to the TE result, it is incomplete because is valid only for a very specific class of 4-velocities. While in a static spacetime, as that used in the TE derivation, one has a unique candidate for the 4-velocity fields necessary to explicit the heat bath's rest frame, in the case of a stationary non-static spacetime (as that in \citep{Buchdahl1949} proof) the rest frame of the bath can be fixed by several 4-velocities fields.

A different approach to derive the TE effect, as well as Tolman-Oppenheimer-Volkoff (TOV) \citep{Tolman1934,Oppenheimer1939},  Klein  related result\citep{Klein}, and in particular the derivation of Einstein's equations from thermodynamics of the self-gravitating gas was attempted by Cocke \citep{Cocke1965}.  He  derived the TOV equation through a maximum entropy principle which was later on extended by \cite{Sorkin1981}. A further generalization to arbitrary perfect fluids was also given by Gao\cite{Gao2011,Gao2012}. Some time after, Roupas\cite{Roupas2013,Roupas2014,Roupas2015,Roupas2018} specified in which thermodynamic ensemble the calculation must be performed thereby recalculating the TOV, TE, and Klein result. It is also worth mentioning that Rovelli and Smerlek\cite{Rovelli2011} also obtained the TE relation by applying the equivalence principle to a property of thermal time.

In this paper, we discuss a general proof of the TE theorem, in a simpler and more general form from that discussed by \citep{Buchdahl1949,Santiago2018}, and similarly we will not use the maximum entropy principle as adopted by many authors\citep{Cocke1965,Sorkin1981,Gao2011,Gao2012,Roupas2013,Roupas2014,Roupas2018}.

As we shall see, using only thermodynamics and general relativity, it is possible to show that under given conditions (null chemical potential) the source of curvature satisfies exactly the TE law. The result is valid for any kind of fluid, not only in the radiative case as originally considered  by Tolman and Ehrenfest.  In the general case of equilibrium (non-null chemical potential), the temperature and the chemical potential and the metric coefficient $g_{00}(x^{r})$ are entertained in such a way that only a definite (position dependent) relation uniting such quantities is obeyed. {Such a result generalize both the TE and Klein relations.}   

The paper is organized as follows. In Sec. II, we obtain the TE relation for a simple fluid regardless the values of its chemical potential, and in Sec. III, we show how the temperature changes due to the TE effect in an isothermal gas, and inside neutron stars. { Finally, it  is closed with a brief summary of the main results.

\section{Thermodynamic States}

As widely known, the thermodynamic states of a relativistic simple
fluid are characterized by three fundamental quantities: (i) an
energy-momentum tensor $T^{{\alpha}{\beta}}$, (ii) a particle
current $N^{\alpha}$, and (iii) an entropy current $S^{\alpha}$.
In addition, the fundamental
equations of motion are expressed by the
conservation law of energy-momentum (${T^{{\alpha\beta}}}_{;\beta}
= 0$) and the number density of particles
(${N^{{\alpha}}}_{;\alpha} = 0$), where the semi-colon denotes
covariant derivative. Moreover, { for a simple fluid, in the absence of classical dissipative
mechanisms (e.g., viscosity, heat flow),} the entropy flux is also conserved quantity
(${S^{{\alpha}}}_{;\alpha}=0$). In an arbitrary hydrodynamic frame
of reference, whose four-velocity obeys $u^\alpha u_\alpha = 1,$
these quantities take the following forms:

\begin {equation}\label{energy}
T^{\alpha \beta} = (\rho + p) u^{\alpha} u^{\beta} - pg^{\alpha
\beta},
\end{equation}
\begin{equation} \label{particle}
N^{\alpha} = nu^{\alpha},
\end{equation}
\begin {equation} \label{entropy}
S^{\alpha} = n\sigma u^{\alpha}.
\end{equation}
In the above relations, the variables $\rho$, $p$, $n$ and $\sigma$ stand respectively for
the energy density, thermostatic pressure, particle number density
and specific entropy (per particle), and are related by the
so-called Gibbs law \cite{D78,SW72,SL02}


\begin{equation} \label{GL}
nTd\sigma = d\rho - {\rho + p \over
n}dn.
\end{equation}
{ The above local expression when combined with the energy conservation law for a perfect fluid ($u_{\alpha}T^{\alpha \beta}_{;\beta} =0$) and the conservation of the particle current ($N^{\alpha}_{;\alpha} =0$),  implies that the $d\sigma/ds$ is conserved along the world lines of the fluid (see, for instance, \cite{SW72}). It means that the flow is isentropic, a result also in accordance with the conserved entropy current (${S^{{\alpha}}}_{;\alpha}=0$).} 

{ Nevertheless, the constancy of $\sigma$ along each world-line,  does not mean  that it is a global constant within the whole volume of the  fluid.  In other words, the constant may vary from world-line to world-line. In particular, for comoving observers  at rest with the volume elements of a {\it static inhomogeneous} simple fluid (the case study here),  $\sigma = \sigma(x^{i})$. It varies from place to place thereby making sense to calculate partial space derivatives in the fluid, and, of course, the same happens with the remaining physical quantities.}
 
{On the other hand}, in the frame defined by (1), $g^{00} = g_{00}^{-1}$, an
observer at rest has 4-velocity ${u^{\alpha}} =
{{\delta}^{\alpha}}_{0}/\sqrt{g_{00}}$. From
$u^{\alpha}u_{\alpha}= 1$ { we also see that $u_{\alpha} = \sqrt
g_{00}{\delta_{\alpha}}^{0}$ while the four-acceleration $a_{\alpha}= u^{\alpha}_{;\beta}u^{\beta}=\Gamma^{0}_{\alpha 0}= -\partial_{\alpha} g_{00}/2g_{00}$. As one may check, by using the above results valid for the static metric (1),  the energy conservation law takes the following form:}

\begin{equation}\label{ECL}
{\partial p  \over \partial x^{i}} + \left[\frac{\rho +
p}{2}\right]g_{00}^{-1}{\partial g_{00} \over \partial x^{i}} =
0,
\end{equation}
or, equivalently,

\begin{equation}\label{GE}
{\partial \ln(\rho + p) \over \partial x^{i}} +  {\partial
\ln{\sqrt g_{00}} \over \partial x^{i}} = \frac{1}{\rho + p}
{\partial\rho \over
\partial x^{i}} \,.
\end{equation}

Now, since the thermodynamic variables are related with the relativistic chemical potential (thermodynamic potential) per particle  by the local Euler expression\cite{Dgroot}

\begin{equation}
T\sigma = \frac{\rho + p}{n} - \mu,
\end{equation}
there are two kind of fluids to be considered, namely those with
and with no chemical potential\footnote{For an ideal relativistic gas, the chemical potential per particle includes the non-relativistic value plus the rest mass-energy contribution, $\mu =\mu_{NR} + mc^{2}$.}.  Let us now discuss each case separately.
\\
\\
(i) $\mu = 0$
\\

{Particles with no chemical potential includes as particular cases the radiation blackbody fluid  (massless photons) as discussed by
Tolman and Ehrenfest and a spin-2 massless field  described by the Fierz-Pauli Lagrangian\cite{FP1939} which coincides with the first order weak field approximation of general relativity (massless gravitons). It should be noticed, however,  that the massless property does not imply a nullified chemical potential. For instance, in special relativity, when $g_{00}=1$ and  $g_{ij}=\delta_{ij}$ in metric (1), a kinetic theoretic approach for an effectively massless (ultra-relativistic) ideal gas yields for the fugacity (in natural units\footnote{$\hbar=k_B=c=1$.}), $exp\large({\mu/T}\large) =\frac{\pi^{2}n}{T^{3}}$.} Note also that since $n \propto T ^{3}$ in this limit,  the fugacity or equivalently, the ratio $\mu/T = constant$.} {Actually, for an ideal noninteracting (and non-quantum) relativistic gas such a property remains valid regardless of the temperature interval\cite{Dgroot}.}

Now, by using that $\rho + p = nT\sigma$, we
first rewrite (\ref{GE}) in the form

\begin{equation}\label{gooT}
{\partial \ln T{\sqrt g_{00}} \over \partial x^{i}} =
\frac{1}{\rho + p} {\partial\rho \over
\partial x^{i}} - \frac{1}{n\sigma}{\partial n\sigma \over
\partial x^{i}},
\end{equation}

or still,

\begin{equation}\label{MU}
{\partial \ln T{\sqrt g_{00}} \over \partial x^{i}} =
\frac{1}{\rho + p}\left[{\partial\rho \over
\partial x^{i}} - \frac{\rho + p}{n}{\partial n \over
\partial x^{i}} - nT{\partial
\sigma \over
\partial x^{i}}\right]\,.
\end{equation}

{As discussed before, in the  inhomogeneous static fluid discussed here, all quantities in the Gibbs law, namely, $\rho$, $\sigma$, $p$, $T$ and $n$ are local functions of the spatial coordinates alone. In this way,  one may think  that the  differentials in (\ref{GL}) are just the differences between infinitely adjacent points of space. 

Notice that all the local thermodynamic properties 
of the fluid can be expressed in terms of $n$ and $T$.
In particular, we have $\sigma = \sigma(n,T)$, 
$\rho = \rho(n,T)$, and $p=p(n,T)$. As a consequence
of the second law of thermodynamics these three
functions have to comply with the differential relation
(7). Consequently, regarding the differentials appearing in (7) 
as the increments associated with neighboring points in 
the inhomogeneou static fluid, it follows that

}

\begin{equation}
nT{\partial \sigma  \over \partial x^{i}} =  {\partial \rho \over
\partial x^{i}}  - \left[\frac{\rho + p}{n}\right]{\partial
n \over \partial x^{i}}.
\end{equation} 


The above equation means \footnote{In this connection see also discussion above and below equation (16).} that the right hand side (RHS) of (\ref{MU}) is
identically null, thereby showing that $ T{\sqrt g_{00}(x^{i})}=\tilde{T}$
as derived by Tolman and Ehrenfest. {\it The present proof is,
however, more general than the original TE theorem since the fluid
is not restricted to blackbody radiation, and, perhaps, more
important, the introduction of a radiation thermometer connecting
two parts of the medium is by no means a necessary device}.
Further, since the equation of state obeyed by the fluid does not
play any role in this approach, such a result strongly suggest
that a general proof including a non-null chemical potential could
in principle be accomplished. This case will be discussed next.
\\
\\
(ii) $\mu \neq 0$
\\

Let us consider again the energy conservation law for the general static configuration.
It is easy to see that equation (\ref{GE}) now leads to the following relation
(compare with Eq. (\ref{MU}))

\begin{equation}\label{MU1}
{\partial \ln T{\sqrt g_{00}} \over \partial x^{i}} =
\frac{1}{\rho + p}\left[{\partial\rho \over
\partial x^{i}} - \frac{\rho + p}{n}{\partial n \over
\partial x^{i}} - nT{\partial
\sigma \over
\partial x^{i}} - \frac{\rho + p}{\sigma + \mu/T}{\partial \mu \over
\partial x^{i}}\right],
\end{equation}
{ and since the first three terms  within the bracket on the {\it r.h.s.} sum zero due to Gibbs law (13), after some algebra the above expression reduces to:}

\begin{equation}\label{MU2}
{\partial \ln T{\sqrt g_{00}} \over \partial x^{i}} +
\frac{\mu}{T\sigma} {\partial \ln \mu{\sqrt g_{00}} \over
\partial x^{i}}=0\,.
\end{equation}

{ It thus follows that a more general relation uniting the pair of thermodynamic variables (T,\,$\mu$) holds regardless of the equation of state satisfied by the medium. It also implies that the original Tolman-Ehrenfest thermodynamic theorem is valid only if the chemical potential vanishes identically or even whether the extended thermodynamic relation for the chemical potential $\mu{\sqrt g_{00}}=constant$ is assumed.  In the general case of equilibrium, the temperature and the chemical potential are entertained in such a way that only a
definite (position dependent) relation uniting both quantities is obeyed. Naturally, the above result is also valid in the particular case of special relativity.}


At this point, it is interesting to comment on the proof
of a related theorem derived long ago by Klein \cite{Klein}. { In his paper, the following shortcut approach was adopted under the same starting conditions, namely: the specific entropy was eliminated from Gibbs-law (\ref{GL}) in order to recover the Gibbs-Duhem relation, $n\sigma dT = dp - nd\mu$. Further, this differential expression was combined with the energy conservation law (\ref{ECL}) thereby obtaining (see Eq. (16) in \cite{Klein})\footnote{Note that in Klein's notation $n \equiv C$, $\mu \equiv \alpha$, $g_{00}\equiv g_{44} = e^{\nu}$, and ${\partial \over {\partial x^{i}}} \equiv \nabla_i$.}. }

\begin{equation}\label{Klein}
n\large[{\partial \mu \over {\partial x^{i}}} - \frac{\mu}{T}\frac{\partial T} {\partial x^{i}}\large] + (\rho + p) \large [\frac {1}{T} \frac {\partial  T} {\partial x^{i}} + {\partial\ln {\sqrt g_{00}} \over
\partial x^{i}}\large ]=0\,.
\end{equation}

{From the above expression it was observed that the temperature and the chemical potential are curiously interrelated. However, Klein took for granted the general validity of the Tolman result (effectively valid only for $\mu = 0$) and concluded that the above expression  (now reduced to the first term) leads to the universal relations: $\mu/T= constant$, and, subsequently (by using the Tolman law again), to the equally celebrated Klein's law:}

\begin{equation}\label{MU3}
\mu{\sqrt g_{00}(x^{i})}=constant\,.
\end{equation}

{ Note that our viewpoint is different by the following reason. We consider that both terms in the brackets are in principle different from zero, unless some extra simplifying condition is assumed (as the general validity of the TE law). As a matter of fact, under more general conditions, the relation $\mu/T$ also does not need to be constant. In general, this happens for an ideal gas of non-interacting particles, a very particular case of the perfect fluid description assumed here. If the fluid obeys a more general equation of state than  the one valid for an ideal gas ($p=nk_B T$), as in the van der Walls case, the ratio $\mu/T$ is different from a constant. This means that the constancy of the fugacity is by no means a general thermodynamic law. Even kinetically, it fails when interactions or even quantum effects are included in the ideal gas description. For instance, for a degenerate relativistic Fermi gas, the exact kinetic result involves special functions, but more enlightening approximate expressions can be obtained for some limits. In particular, in the almost complete degeneracy regime, $T/T_F << 1$  where $T_F$ is the Fermi temperature,  the chemical potential can be written as:

\begin{equation}
\mu = E_F\large[ 1 - {\pi^{2} \over 12} \large(\frac{T}{T_F}\large)^{2} + ...\large] + mc^2,
\end{equation}
where $E_F$ is the Fermi energy and $T_F=E_F/k_B$ (see, for instance, \cite{FG}).

The above  considerations lead us to conclude that the standard TE and Klein's thermodynamic relations laws are generically valid only for the restricted class of perfect fluids satisfying the relation $\mu/T = constant$. Basically, noninteracting (ideal) relativistic gas when quantum effects are not considered [in this connection see also comment on fugacity just above equation (11)].

\section{The case of mixtures}

{In the present work all calculations are restricted to one-component  self-gravitating  relativistic simple fluids, 
either with vanishing  or with finite chemical potentials. Naturally, such results cannot be generically  applied to mixtures 
without careful further considerations on the thermodynamic variables, as well as 
on the specific properties of each of the component comprising the mixture. 

In general, even when locally the temperature is the same for each substance in the mixture, there are $P$ different chemical potentials 
$\mu_i$, $i=1,2,...P$, for $P$  independent substances. This means that the Gibbs,  Euler and the remaining 
relations, including the energy-momentum tensor must now be written as a sum involving all components. 

In the case of a static mixture of matter and radiation, for instance, the results may also depend on the 
value of local temperature  within the system. The local equilibrium radiation has null chemical potential 
and the same happens with the material component when the mass of the particles (in natural units) is much smaller than the temperature ($T >> m$). 

Naturally, our results may be extended to more complex situations by taking into account the proper extensions 
of the basic equations. Nevertheless, a detailed treatment involving several substances is out of the scope of 
the present paper and will be discussed in a forthcoming communication.} 

\section{Tolman-Ehrenfest law in ideal gases and neutron stars}

The previous calculations showed a general proof of the Tolman-Ehrenfest effect. In this section we want to show how 
this effect manifest itself in thermal equilibrium ideal gases (isothermal spheres, in spherical, bounded, static configurations), and in neutron stars.
To this goal, we will follow \cite{Roupas2014,Roupas2015}.

\subsection{Ideal gases}

In GR the equation of state of the relativistic ideal gas, in a sphere of radius $R$, can be expressed as \cite{Israel1963,Roupas2014}
\begin{equation}\label{eq:eos}
P(r) = \frac{1}{T_{\rm inv}(r) \left[1+L(T_{\rm inv}(r)) \right]}\rho(r) c^2,
\end{equation}
where $T_{\rm inv}= \frac{mc^2}{kT}$ and $L(T_{\rm inv})$ is given by
\begin{equation}\label{eq:F}
L(T_{\rm inv}) = \frac{K_1(T_{\rm inv})}{K_2(T_{\rm inv})} + \frac{3}{T_{\rm inv}} - 1\,,
\end{equation}
and $K_n$ are the modified Bessel functions:
\begin{equation}\label{eq:bessel}
	K_n(T_{\rm inv}) = \int_0^{\infty} e^{-T_{\rm inv} \cosh\theta}\cosh(n\theta)d\theta,	
\end{equation}
The thermal and dynamic equilibrium is given by the following four equations: TOV equation \cite{Tolman1939,Oppenheimer1939}
\begin{equation}\label{eq:TOV}
({\bf i}) \hspace{0.5cm} \frac{d P}{dr} = - \left({\frac{P}{c^2}} + \rho\right) {\left(\frac{G{M}}{r^2} + 4\pi G \frac{P}{c^2} r\right) \left(1 - \frac{2G{M}}{rc^2} \right)^{-1} },
\end{equation}
the mass equation
\begin{equation}\label{eq:mass_d}
({\bf ii}) \hspace{0.5cm} \frac{d {M}}{dr} = 4\pi \rho r^2,	
\end{equation}
where $M$ is the total mass (namely the sum of the rest mass,
the thermal energy, and gravitational field's energy)
the Tolman-Ehrenfest relation \cite{Tolman1930,Tolman1930a}
\begin{equation}
({\bf iii}) \hspace{0.5cm} T(r) \sqrt{g_{00}}=constant,
\end{equation}
that can be written in  differential form as \citep{Gao2012,Roup}
\begin{equation}\label{eq:tolman}
({\bf iv})\,\,\,\frac{d T_{\rm inv}}{dr} = -\frac{T_{\rm inv}}{P+\rho c^2} \frac{dP}{dr}, 	
\end{equation}
and to close the system we use Eq. \ref{eq:eos}. We then solve Eqs. \ref{eq:TOV}, \ref{eq:mass_d}, \ref{eq:tolman}, and \ref{eq:eos}.

\begin{figure*}[ht]
\begin{center}
\includegraphics[scale=0.4]{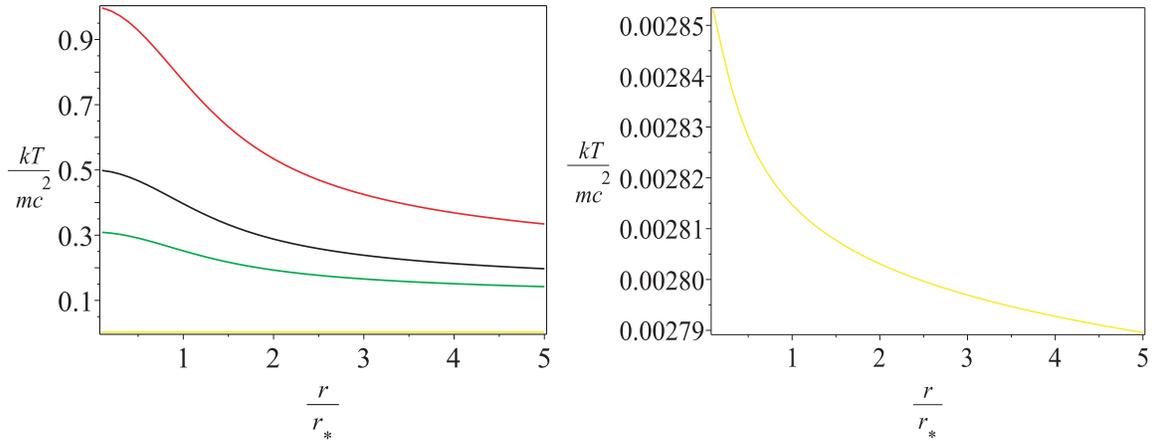}
\caption{The Tolman-Ehrenfest effect. The plot shows the proper temperature for different central temperature, $T_0=\frac{m c^2}{k b_0}$. Left panel: the red line represents the solution in the case $b_0=1$, while the black line, green, and yellow lines the case $b_0=2$, $b_0=3.23$, and $b_0=350$ respectively.
The quantity $b_0= \frac{mc^2}{k T_0}$, is fundamentally the inverse of the temperature, while the normalizing factor on the $x$ axis, $r_*=1/\sqrt{4\pi G\rho_0 m k T_0}$, where $\rho_0$ is the central density. Right panel: the case $b_0=350$ using a different scale.
%
	\label{fig:Tolman_Ehrenfest}}
\end{center}
\end{figure*}

The system of differential equation must be solved with the initial conditions: $\rho(0)=\rho_0$, $T_{\rm inv}(0)=T_0$, $M(0)=0$, with $r \in [0, R]$. The equilibrium equations can be expressed in adimentional form as shown in Eqs. (58-60) of \cite{Roupas2014}, and the relative initial conditions.

In Fig. 1, we plot the result of the integration, namely the Tolman-Ehrenfest effect: the proper temperature versus the radius. The quantity $b_0= \frac{mc^2}{k T_0}$, is fundamentally the inverse of the temperature, while the normalizing factor on $x$ axis, $r_*=1/\sqrt{4\pi G\rho_0 m k T_0}$, where $\rho_0$ is the central density. The red line represents the solution in the case $b_0=1$, while the black line, green, and yellow lines the case $b_0=2$, $b_0=3.23$, and $b_0=350$ respectively.
The plot shows that the larger is the central temperature the larger is the TE effect, and that the temperature gradient decreases with distance from the system center.
One can also define a rest mass as
\begin{equation}
M_{\rm rest}= \int_0^R \frac{m n(r)}{\sqrt{1-\frac{2GM}{rc^2}}} 4 \pi r^2 dr,
\end{equation}
where $n(r)$ is the particle density, that can be expressed as \citep{Roupas2018}
\begin{equation}
n(r)=n_0 \frac{K_2(T_{\rm inv})}{K_2(T_{\rm inv}(0))} \frac{T_{\rm inv}(0)}{T_{\rm inv}},
\end{equation}
which is given, for the four cases considered, by $0.78 M_S$, $0.66 M_S$, $0.56 M_S$, and $0.011 M_S$, with $M_S=\frac{R c^2}{2 G}$.

%
%

\begin{figure}[ht]
\begin{center}
\includegraphics[scale=0.4]{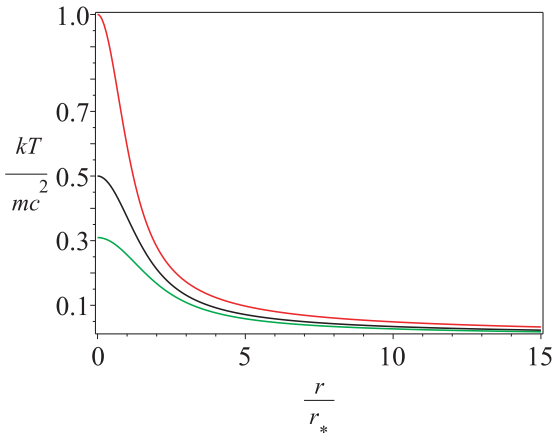}
\caption{The Tolman-Ehrenfest effect. The red line represents the solution in the case $b_0=1$, while the black line and green line the case $b_0=2$, and $b_0=3.23$ respectively.
The quantity $b_0= \frac{mc^2}{k T_0}$, is fundamentally the inverse of the temperature, while the normalizing factor on $x$ axis, $r_*=1/\sqrt{4\pi G\rho_0 m k T_0}$, where $\rho_0$ is the central density.
%
	\label{fig:Tolman_Ehrenfest}}
\end{center}
\end{figure}

\subsection{Neutron stars}

In this section, we discuss how the TE effect changes the inner temperature of NSs.

NSs are important laboratories in which extreme conditions of matter can be studied. Their EoS and composition are unknown at density larger than $\rho_0= 2.8 \times 10^{14} \rm g/cm^3$ \cite{Lattimer2000}, and different models predict different composition and EoS. Cooling is a powerful method to have insight on the inner structure of NSs \cite{Page1998}.

NSs are very hot immediately after the supernovae explosion. Their temperatures is $T \simeq 10^{11} \rm K$
large gradient temperatures are present. In a conduction timescale, the heat flows inward, generating a cooling wave propagation from the NS center to its surface.
In usual calculations \cite{Prakash1997}, the TE effect is not taken into account, while in others, \cite{Gnedin2001}, it is claimed that the NS become isothermal in times of the order of 50-100 yrs.

In reality,  neutron stars  are not isothermal at all. In their Figs. 6-7, \cite{Gnedin2001} are not plotting the local temperature $T(r)$, but the so called red-shifted temperature $\tilde{T}= T(r,t) \exp{\phi(r)}$, where $\phi(r)$ is the potential. Locally the temperature changes from one point to the other, and the system is not isothermal.
%
%
%

In order to find the gradients of temperature in the NS due to the TE effect, we will solve Eqs. (\ref{eq:TOV},\ref{eq:mass_d}, \ref{eq:tolman}), coupled with an EoS. The EoS that we use is that of \cite{Kurkela2014} constrained by using info coming from the high-density limit from perturbative QCD, from low-energy nuclear physics, and pulsars data. For $\rho < 3.3 \times 10^3 \rm g/cm^3$ the EoS used is that of \cite{Harrison1965}, while for the inner and outer crust we use \cite{Negele1973,Ruester2006}. Solving the quoted equations, we get the results plotted in Fig. 2, in which the colors correspond to that of Fig. 1. In Fig. 2 the temperature falls in a steeper way than in Fig. 1, however the temperature gradient is always present, as predicted by the TE effect. The changes from 5 km to the crust of the NS becomes very small and the behavior tend to become more isothermal.

\section{Conclusions}

In the present paper, we advance a proof of the TE theorem that is more general 
than the ones proposed in \cite{TE30,Tolman34,Klein}, or than proofs 
based on a maximum entropy principle, such as those reported in \cite{Sorkin1981}
or in \cite{Gao2011,Gao2012}. In our analysis the idea of a radiation thermometer 
is not necessary. The derivation presented here is as independent as possible 
of the properties of specific media. In that sense, it has a generality 
con-substantial with the robustness that a fundamental thermodynamic principle 
is expected to have. Other derivations of the TE law, such as the simplified 
ones based on gravitational redshift \cite{TP}, in spite of their considerable
heuristic and pedagogical values, lack the above mentioned kind of generality.

The main aspects of our proof are: (i) the fluid is not restricted to blackbody radiation, and the result within the approach followed here can be naturally extended for fluid mixtures (ii) the radiation thermometer connecting two parts of the medium is not a necessary device, and (iii) the proof is independent from the equation of state and, as such, it was also possible to provide a more general  proof including a  non-null chemical potential. Finally,  we have solved the thermal and dynamic equilibrium equations (TOV, mass equation, TE relation) to find the relation between the temperature and radius, in the case of an isothermal gas, and neutron stars. The result shows that the temperature gradients are larger for larger central temperatures, and are larger for the NS case with respect to the isothermal gas case.

\vskip 0.5cm {\bf Acknowledgments:} One of us (JASL) is grateful to
the warm hospitality at the Physics Department of the University of Pretoria (South Africa). Partial support from CNPQ, CAPES (PROCAD) and FAPESP (LLAMA project) is also acknowledged.

\end{document}